\documentclass[showpacs,prl,twocolumn]{revtex4}

\usepackage{epsfig}
\usepackage{amsmath}
\usepackage{graphicx}
\usepackage{dcolumn}
\usepackage{amsmath}
\usepackage{latexsym}
\usepackage{color}
\begin{document}

\title{\textbf{A Superconducting Gap in an Insulator}}
\author{D. Sherman$^{1}$}
\altaffiliation{Laboratory for superconductivity and optical spectroscopy, Ariel University center, Ariel 40700, Israel.}
\author{G. Kopnov$^{2}$, D. Shahar$^{2}$ and A. Frydman$^{1}$}
\affiliation{$^{1}$ The Department of Physics, Bar Ilan University, Ramat Gan 52900,
Israel\\
$^{2}$ Department of Condensed Matter Physics, Weizmann Institute of Science, Rehovot 76100, Israel}
\date{\today}

\begin{abstract}
We present tunneling spectroscopy and transport measurements on disordered indium oxide films that reveal the existence of a superconducting gap in an insulating state. Two films on both sides of the disorder induced superconductor to insulator transition (SIT) show the same energy gap scale at low temperatures. This energy gap persists up to relatively high magnetic fields and is observed across the magnetoresistance peak typical of disordered superconductors. The results provide useful information for understanding the nature of the insulating state in the disorder induced SIT.
\end{abstract}

\pacs{74.55.+v, 74.62.En, 74.81.Bd}
\maketitle

\date{\today}

Increasing the disorder of a metallic system causes the localization of its electronic wave function. If the ground state of the system is a superconductor, increasing the disorder leads to a transition from a superconductor to an insulator (SIT). This transition has gained increasing attention lately due to the experimental observations of a number of dramatic features near the SIT such as simple activated temperature dependence of the resistance on the insulating side \cite{shahar_ovadyahu}, a large peak in the magneto-resistance \cite{hebard,gant,Sambandamurthy,Steiner,BaturinaC}, peculiar I-V characteristics \cite{iv1,iv2} and traces of superconductivity at temperatures above $T_C$ \cite{sacepe0,sacepe2, armitage,pratap}. Another reason for renewed interest in this field is that the SIT may be a basic realization of a quantum phase transition which occurs at T = 0 as a function of disorder or magnetic field and is driven by quantum rather than thermal fluctuations. Despite this growing interest, the mechanism of the SIT and, in particular, the nature of the insulating state are not understood. Recently a few indications for the presence of electronic pairs in the insulator have been reported \cite{sacepe2,hollen,shachaf}, inspiring further theoretical effort. Two general ideas have been put forward to try to explain some of the observations. One relies on the role of the disorder in generating inhomogeneity in structurally homogeneous samples \cite{kowal,efrat,dubi,imry,sacepe1}. Within this framework, the crossover from insulating to superconducting behavior occurs when the Josephson-coupling between superconducting islands succeeds in forming a percolation path throughout the system. A second scenario invokes the existence of uncorrelated pre-formed electron pairs which do not constitute a condensate  but are characterized by an energy gap that is associated with the pair binding energy \cite{feigel}. Other models adapt concepts from both pictures and suggest that the insulating film is composed of small superconducting islands that are uncorrelated and are too small to sustain bulk superconductivity \cite{nandini, nandini2}. Clearly, additional experimental results are required to help shed light on this "supercoducting insulator".

In this letter we present an experimental study of the DOS and corresponding transport characteristics of two disordered films on both sides of the transition. These measurements show that a similar energy gap exists in both the superconducting and the insulating states. We present the dependence of this gap on disorder and magnetic field and discuss the possible implications of these experimental results towards the understanding of superconductivity in highly disordered films.

For measuring the DOS of the films we fabricated tunnel junctions in the following way: A 30 nm Al stripe was thermally evaporated on a Si/SiO substrate and was allowed to oxidize for a few hours in ambient conditions.  Subsequently, a 31 nm thick indium oxide (InO) stripe was e-gun evaporated perpendicular to the Al stripe, thus forming a planar tunnel junction with barrier dimensions of 1mm*1mm.  In order to produce InO films with different disorder, dry oxygen was injected into the evaporation chamber at different partial oxygen pressure \cite{zvi_86}. This resulted in amorphous yet structurally homogeneous films with different degrees of disorder characterized by their sheet resistance $R_{\Box}$.

The results presented in this letter were obtained on two amorphous InO films; one exhibiting superconducting transport (sample S) and the other exhibiting insulating behavior (sample I). Fig. 1 shows the resistance versus temperature of both films. It is seen that while sample S shows a clear superconducting transition with $T_C \sim 3K$, the resistance of sample I increases rapidly with lowering T down to our base temperature of $25mK$, thus showing insulating properties.  A fit to Arrhenius law:
\begin{eqnarray}
R=R_{0}exp(T_{0}/T).
\end{eqnarray}
is shown in Fig 1. From this fit one yields an activation temperature $T_{0} = 0.4K$ \cite{rem0}, and $R_{0} = 11.5k\Omega$ which is larger than the quantum resistance for pairs, $1/G_0=h/2e^{2} = 6.45k\Omega$.

\begin{figure}[h]
\vspace{0cm}
    \centering
    \label{rt1}
    \includegraphics[width=\columnwidth,keepaspectratio=true]{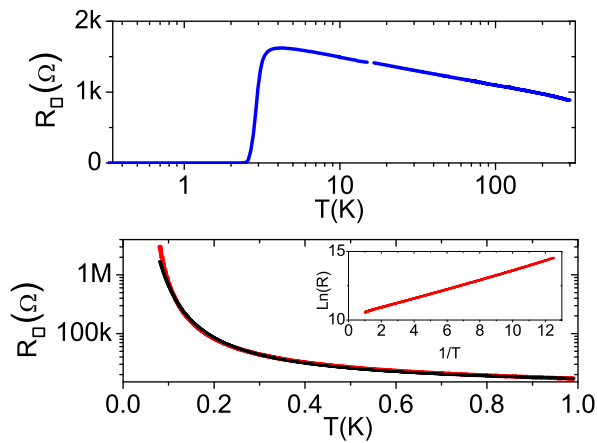}
    \vspace{-0.5cm}
    \caption{(color online). Top: Temperature dependence of $R_{\Box}$ for sample S. The corresponding AlO barrier resistance is $1M\Omega$. Bottom: Temperature dependence of $R_{\Box}$ for sample I (heavy red line) and a fit to Eq. 1 (light black line) from which $T_0$ and $R_0$ are extracted. The corresponding AlO barrier resistance is $0.7M\Omega$. Inset: $ln(R)$ versus $1/T$ for sample I.}
    \vspace{-0.6cm}
\end{figure}

Despite this clear difference in the transport properties, the tunneling spectra of the two films appears to be surprisingly similar. Fig. 2 depicts dI/dV versus V curves for both samples. All tunneling measurements presented here were performed by standard lock-in techniques while making sure that the junction resistance was at least an order of magnitude larger than the InO sheet resistance so that the film could be regarded as an equi-potential electrode. Since these are disordered films one has to take into account that electronic interactions cause a suppression of the density of states (DOS) at low energies either due to the Altsuler Aronov \cite{AA} zero bias anomaly (ZBA)  mechanism for weak disorder or due to the coulomb gap \cite{ES} for high disordered systems.  In order to isolate the superconducting contribution to the DOS we normalized the curves by the tunneling spectra of the films taken at a magnetic field, H, of 11T. The justification for this procedure relies on the assumptions that superconductivity is fully suppressed at $H=11T$, and that the magnetic field hardly affects the normal state DOS. Indeed, the dI/dV-V curves of both samples exhibit very small magnetic field dependence at energies larger than the superconductive gap, and for $H \geq 7.5T$ (where superconductivity is suppressed) the curves are practically indistinguishable.

The dI/dV-V curves shown in Fig. 2 demonstrate that a superconducting gap exists in both samples, despite the fact that sample I clearly shows insulating transport behavior. In order to extract the value of the superconducting gap, $\Delta$, we fit these curves to the BSC expression modified by a broadening parameter $\Gamma$ that accounts for the finite scattering time of the superconducting quasi-particles \cite{dynes}:

\begin{equation}
\label{eq_BCS} N_S(E)/N_N(E)=\Re\left\{(E-i\Gamma)/     [(E-i\Gamma)^{2}-\Delta^{2}]^{1/2}        \right\}
\end{equation}

The fits depicted in Fig. 2 show that the experimental curves of both films deviate from the BSC theoretical predictions. In particular, the so called "coherence peaks" at the gap edges are suppressed in our disordered films. Nevertheless, the best fits yield $\Delta = 0.7mV$ \emph{for both films } \cite{rem1}. The only observed differences between the samples are a smooth suppression of the coherence peaks and an increase of the sub-gap conductance.

\begin{figure}[h]
\includegraphics[width=0.5\textwidth]{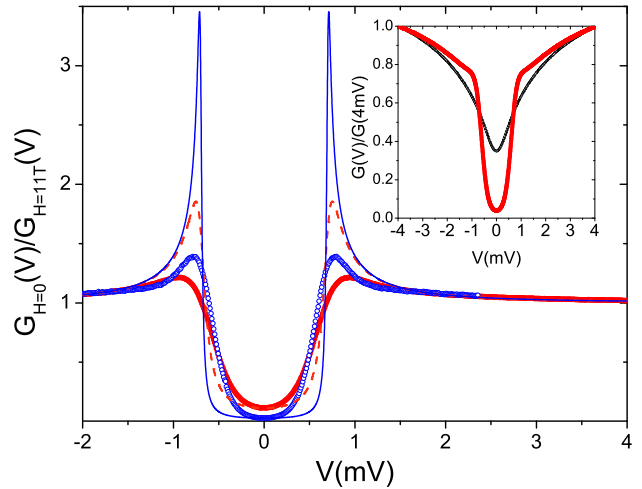}
\label{dos}
 \vspace{-0.5cm}
\caption{(color online). Normalized tunneling density of state obtained at 1K for sample I (full red circles) and sample S (empty blue circles). The solid lines are fits to the BSC expression of Eq. 2 for sample I (dashed red line) and sample S (solid blue line). Inset: Raw dI/dV versus V data for sample I at H=0 (heavy red line) and H=11T (light black line).}.
  \vspace{-0.2cm}
\end{figure}

The fact that the superconductive gap persists into the insulating side of the SIT was experimentally implied in the past. Scanning tunneling microscopy measurements of the local DOS were performed on disordered TiN \cite{sacepe1, sacepe0} and InO \cite{sacepe2} superconductors with different degrees of disorder. In both cases it was found that $\Delta$ does not decrease with disorder as would be expected from the decrease of $T_C$ extracted from the transport measurements. In the TiN films $\Delta$ decreased with increasing disorder, however  $\frac{T_C}{\Delta}$ was found to decrease as the films approached the SIT. In the InO samples the average $\Delta$ showed no clear dependence on disorder. In both cases an extrapolation would predict a finite gap in the insulator. Our measurements confirm this trend since we observe the same gap magnitude on the two sides of the transition. This observation is in accordance with a recent theoretical work \cite{nandini2} which predicts that $\Delta$ is expected to remain unchanged through the SIT and may even grow in magnitude deep in the insulator.

\begin{figure}[h]
\includegraphics[width=0.5\textwidth]{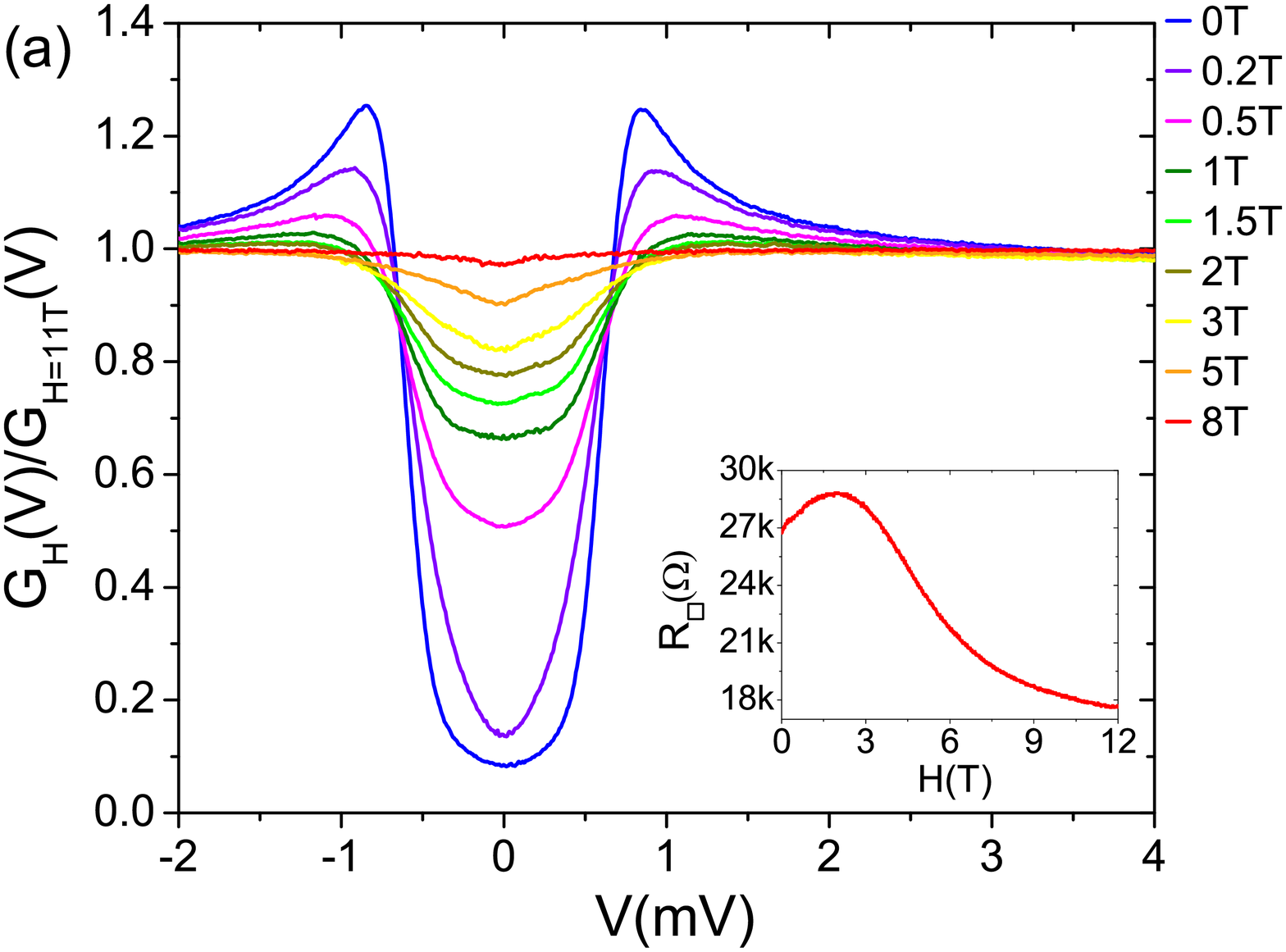}
\includegraphics[width=0.5\textwidth]{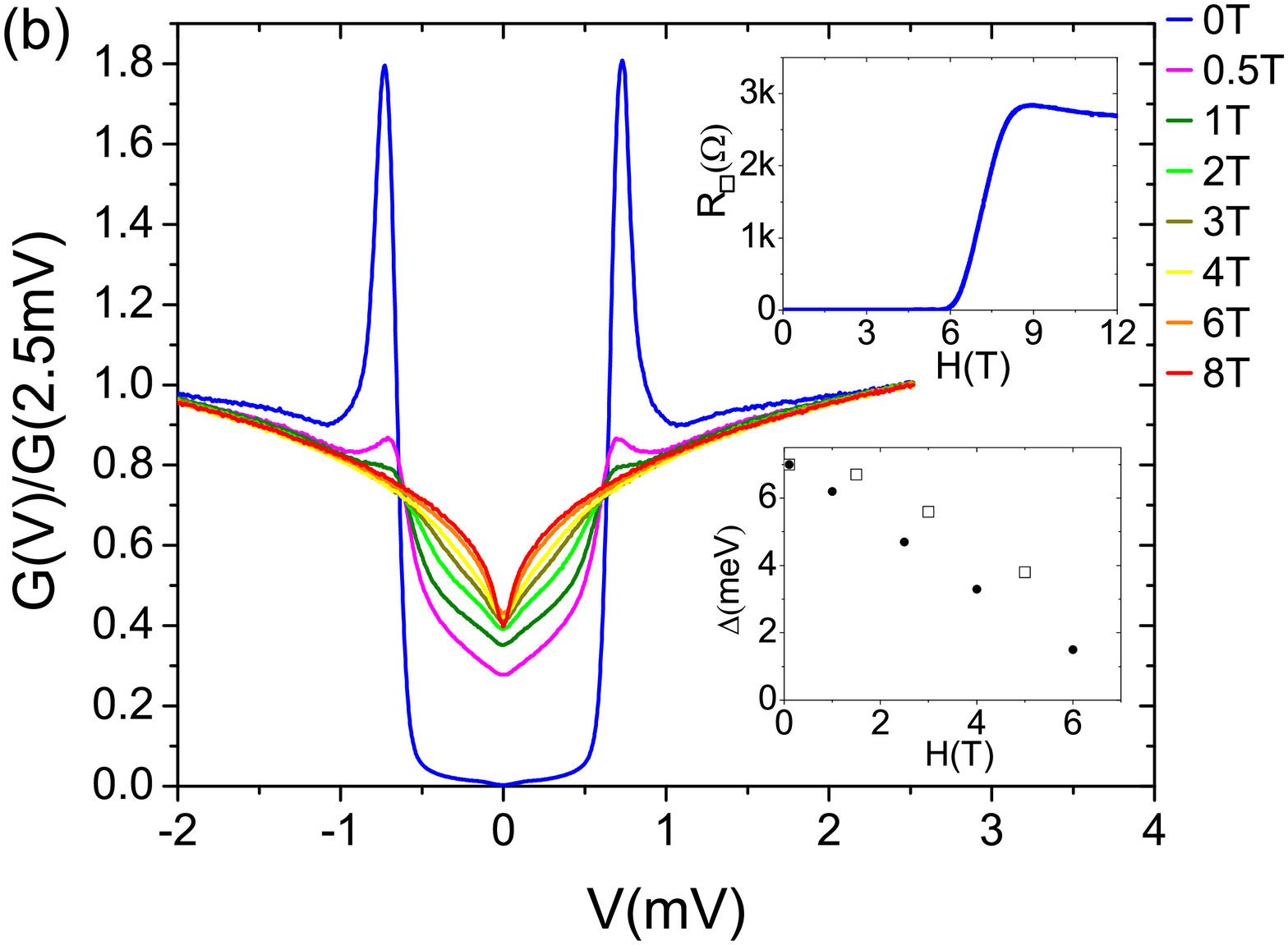}
\label{dos_h}
\caption{ (color online). (a) Normalized tunneling density of state obtained at T=0.5K and at different magnetic fields for sample I. inset: the corresponding magnetoresistance. (b) Normalized tunneling density of state obtained at 0.1K and at different magnetic fields for sample S. upper inset: The corresponding magnetoresistance for sample S. lower inset: $\Delta$ versus $H$ for sample I (empty black squares) and sample S (full black circles), obtained from a fit to the BCS expression of eq.(2).}
  \vspace{-0.4cm}
\end{figure}

As may be expected, the magnetic field, H, has a significant effect on $\Delta$. Fig 3 shows the dependence of dI/dV versus V curves on H for samples S and I. It is seen that for both samples $\Delta$ decreases monotonically with H and is wiped out altogether at $H \sim 7.5T$. It is interesting to relate these results to the magnetoresistance of the samples. Fig 3 shows that both samples exhibit a magnetoresistance peak similar to that reported in the past \cite{hebard,gant,Sambandamurthy,Steiner,BaturinaC}. The magnetic field at which the peak occurs, $H_P$, is lower in sample I (1.7T at 1K compared to 6.5T for sample S). This is consistent with the trend seen in superconducting samples where $H_P$ was found to decrease with increasing disorder \cite{baturina}. Comparing the tunneling and the magnetoresistance results reveals the fact that while in sample S the gap is wiped out at fields lower than $H_P$, in sample I $\Delta$ persists up to fields that are larger than $H_P$. It appears, therefore, that the DOS is unaffected by the magnetoresistance details, and particularly by the resistance peak, and only depends on the value of H. This demonstrates, again, that while the resistance of the two samples may behave differently, their DOS is very similar. It has been suggested \cite{dubi} that the MR peak is due to the formation of superconducting islands with sizes that depends on the magnetic field. In this picture, $H_P$ corresponds to the magnetic field at which it is preferable for the current to flow through normal regions rather than through the superconducting islands. Our results are consistent with this picture since it implies that the resistance would be sensitive to $H_P$ but the DOS would not be affected by it.

 The results presented above strongly suggest that the disorder driven superconductor to insulator transition is of a geometrical percolation nature. The fact that the DOS changes smoothly across the transition and that the order parameter is hardly affected by it is consistent with a model of superconducting islands of which the effective sizes gradually decreases with disorder. The observation of coherence peaks in the insulator suggests that superconductive grains that sustain a coherent condensate are present in this phase. We envision that the average size of these islands decreases with increasing disorder so that a larger percentage of islands falls below the Anderson limit for superconductivity. This causes the coherence peaks to shrink with disorder. Further increasing of the disorder is expected to cause the coherence peaks to vanish altogether, however, it is very difficult to perform tunneling junction experiments in this regime of high film resistance.

\begin{figure}
\vspace{-0.4cm}
\includegraphics[width=0.5\textwidth]{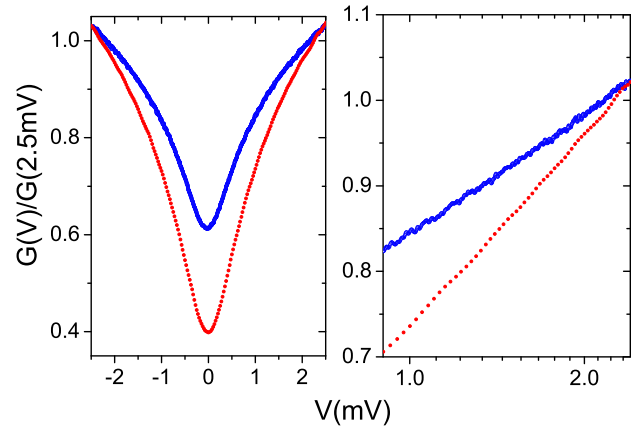}
\label{normal}
\vspace{-0.2cm}
\caption{ (color online). Left: Tunneling density of state versus V (linear scale (Left) and logarithmic scale (right)) obtained at 1K and 11T for sample I (full red circles) and for sample S (blue line). }
  \vspace{-0.4cm}
\end{figure}

This "granular" model is further supported by the results of the DOS in the normal state. As noted, for $H > 7.5T$ the superconductive gap is fully suppressed. At these fields dI/dV is found to be proportional to ln(V) (see Fig. 4). This behavior is consistent with the zero bias anomaly model for weak disorder \cite{AA}. Due to the nearby Al electrode the long range Coulomb interactions are screened. Hence for low voltage and temperature, such that the thickness of the sample is smaller than the thermal length, the DOS is expected to follow the screened 2D expression:

\begin{eqnarray}
    \nu(\epsilon, T) -\nu (\infty)=-\frac{\nu (\infty)}{4  \pi^{2} \hbar g}\ln\frac{2\kappa b}{k_{2}}\ln\frac{\tilde{\varepsilon}\tau}{\hbar}
         \end{eqnarray}

where $\tilde{\varepsilon}=max(V,T)$, $\nu (\infty)$ is the DOS at high energies, $\tau$ is the inelastic relaxation time, $k$ is the dielectric function, $\kappa$ is the inverse screening length, $b$ is the barrier thickness and g is the dimensional conductance $G/G_{0}$. Since the slopes of the dI/dV versus ln(V) curves are proportional to 1/g, our results enable us to extract the conductance ratio between the two films. This yields $\frac{g_S}{g_I}=1.6$. We stress that this ratio does not depend on the exact model for ZBA as long as the 1/g dependence is valid. On the other hand, the conductance ratio extracted from the transport at H=11T gives $\frac{g_S}{g_I}=11$. This dramatic difference can be interpreted as an indication for granularity in the film. While the transport g is sensitive to the current carrying network, tunneling may take place into the metallic regions which have a relatively high conductance even in the insulating phase. In this picture the difference between the DOS in the metal and in the insulator may be very small. 

In conclusion, We have provided direct evidence for the existence of a superconductive gap in an insulator state. The gap amplitude is hardly affected by the dramatic change in the global transport properties. The results strongly suggest that homogeneously disordered superconducting films near the SIT contain superconducting islands with bulk properties that are similar on both sides of the transition. The DOS is not influenced by the SIT nor by the magnetoresistance peak indicating that the insulator contains regions of finite order parameter amplitude that are similar to those in the superconductor.

We are grateful for the useful discussions with D.B. Gutman, B. Sac\'{e}p\'{e}, E. Shimshoni and N. Trivedi. This research was supported by the US Israel binational fund (grant No. 2008299)


\begin{references}
\bibitem{rem0} Noth that $T_0$ is much smaller than the energy gap of this film as shown below.
\bibitem{rem1} Practically, the dI/dV curves represent a convolution of the DOS of the indium oxide and the Al electrodes. However, the Al gap energy is less than a fifth of the indium oxide gap. In addition, the Al was prepared such that its $T_C$ was $0.4K$ in Sample I, and $0.9K$ in sample S. The presented measurements are at higher temperatures, except for the curves in Fig. 3(b), in which the effect of the Al gap can be seen in a log representation in the 0T curve.
\end{references}
\end{document}